# Large Scale Velocity Fields Present and Future: Making Sense of the data†

**Hume A. Feldman**‡

Department of Physics, Princeton University
Princeton, NJ 08544

## Abstract

The large scale velocity field was sampled recently by two independent methods: the supernovae type Ia light curve shapes (Riess, Press & Kirshner) and the Abell Cluster Catalog brightest cluster galaxy metric luminosity (Lauer & Postman). The results of these investigations seem to be contradictory. We analyze these samples, compare them and investigate whether standard structure formation models and other deep surveys are compatible with them. We also make suggestions as to how to improve the samples so we can actually resolve the bulk flow vectors. We show that although these two samples seem to cover the same volume, their window functions are sufficiently different so that they are only weakly correlated. Further, since both samples are sparse, they are noise dominated and in order to improve the signal to noise they need to either increase their sample size (RPK) or decrease the measurement errors (LP) significantly.

## I. Introduction

The large scale velocity field probes the matter distribution in the Universe directly and not merely the light distribution as redshift surveys do. However, to measure the velocity field one needs to make accurate distance measurements which is quite difficult. The errors in these estimates are typically some fraction of the redshift of the sample points. In the case of distant objects, the errors are larger than the velocity vector one wants to measure. This is partially rectified by measuring the lowest moment of the velocity field, namely the bulk flow, where the noise in the measurement is reduced by the square root of the number of objects in the sample.

There were two recent efforts to measure the bulk velocities of large volumes. One was done by Lauer & Postman (1994, hereafter LP) who used as their distance measure the brightest cluster galaxy (BCG) metric luminousities of the complete sample of 119 Abell clusters up to 15,000 km/s (median redshift ≈ 7,500 km/s). Their distance estimates are accurate to about 17% of the redshift.

They found that the Abell cluster inertial frame (ACIF) exhibits a bulk velocity of $\approx 700$ km/s with respect to the cosmic microwave background (CMB) rest frame. More recently Riess, Press & Kirshner (1995, hereafter RPK) developed a distance measurement based on the light curve shapes of 13 type Ia supernovae (SNIa), with distance errors reported to be about 5% of the redshift (median redshift $\approx 5,500$ km/s). RPK found that the local group's velocity with respect to the frame defined by their sample was consistent with that of the CMB rest frame. In recent analyses (Feldman & Watkins 1994, Strauss *et al.* 1995, Jaffe & Kaiser 1995, Watkins & Feldman 1995) it was shown that power spectra from popular structure formation models (CDM, CHDM and PIB) as well as the IRAS–QDOT redshift survey are inconsistent with the LP measurement to $2-3\sigma$ level, whereas they are highly consistent with the RPK result to better than $1\sigma$. Further, the RPK and LP results seem to be inconsistent with each other at a high confidence level ($\sim 99\%$). We will see below that this high level of disagreement is not surprising given standard models of structure formation. For these models, both measurements are expected to be noise dominated and may have large signal because of incomplete cancellation of small scale flows. This, in turn leads to the samples being nearly uncorrelated, although they *seem* to probe similar volumes.

Here we explore the implications of both measurements, we calculate the $\chi^2$ statistic for the RPK and LP results taken independently and together for a variety of power spectra. We also present the expected correlations between the two results. Finally we discuss the expected sensitivity of the samples for determining the underlying bulk flow of the volumes in which they are embedded. We examine how this sensitivity will improve by increasing the sample size in RPK case or decreasing the measurement errors in the LP case.

## II. Analysis

The analysis here follows closely the analyses presented in two previous papers (Feldman & Watkins 1994, Watkins & Feldman 1995). Given a bulk flow vector $U_i$, the covariance matrix for the estimated bulk flow of a sample of galaxies is the sum of two statistically independent parts,

$$R_{ij} \equiv \langle U_i U_j \rangle = R_{ij}^{(v)} + R_{ij}^{(\varepsilon)} \qquad (1)$$

The first part arising from the sampling of the underlying velocity field and the second arising due to the noise in the distance estimates.

$$U_i = A_{ij}^{-1} \sum_n \frac{\hat{r}_{n,j} S_n}{\sigma_n^2 + \sigma_*^2} \qquad (2)$$

$\hat{r}_{n,j}$ is the $j$th component of the unit vector of the $n$th galaxy, $\sigma_*$ is the dispersion in the line–of–sight velocity due to random velocities (which we take to be 400 km/s), $\sigma_n$ is the estimated uncertainty in the line–of–sight peculiar velocity.

$$A_{ij} = \sum_n \frac{\hat{r}_{n,i} \hat{r}_{n,j}}{\sigma_n^2 + \sigma_*^2} \quad (3)$$

The velocity part of the covariance matrix is the convolution of the squared tensor window function and the velocity power spectrum

$$R_{ij}^{(v)} = \int \frac{d^3k}{(2\pi)^3} \mathcal{W}_{ij}^2(\vec{k}) P_v(\vec{k}) \quad (4)$$

where the velocity power spectrum is defined in terms of the power spectrum:

$$P_v(k) \equiv \langle |v(\vec{k})|^2 \rangle = \frac{H^2 a^2}{k^2} P(k) . \quad (5)$$

and the squared tensor window function is

$$\mathcal{W}_{ij}^2(\vec{k}) = W_{il}(\vec{k}) W_{jm}(\vec{k}) \hat{k}_l \hat{k}_m \quad (6)$$

where $W_{ij}$ is the tensor window function for the sample in Fourier space.

$$W_{ij}(\vec{k}) = A_{im}^{-1} \sum_n \frac{\hat{r}_{n,m} \hat{r}_{n,j}}{\sigma_n^2 + \sigma_*^2} e^{i \vec{k} \cdot \vec{r}_n} . \quad (7)$$

We define a $\chi^2$ statistic for the three degrees of freedom of the measured bulk flow vector $\vec{V}$ where $V_i$ is the $i$th component of $\vec{V}$.

$$\chi_v^2 \equiv V_i R_{ij}^{-1} V_j . \quad (8)$$

$R_{ij}$ can also be used to calculate an expectation value for the magnitude of the bulk flow $\Lambda$.

To get an idea of how much correlation we expect between $\vec{U}^{LP}$ and $\vec{U}^{RPK}$ for a given power spectrum by calculating the normalized expectation value for their dot product, which should be close to 1 for highly correlated vectors, zero for vectors that are completely uncorrelated, and $-1$ if there is a high degree of anti-correlation.

$$\mathcal{C} = \frac{\langle U_i^{LP} U_i^{RPK} \rangle}{(\langle U_l^{LP} U_l^{LP} \rangle \langle U_m^{RPK} U_m^{RPK} \rangle)^{\frac{1}{2}}} . \quad (9)$$

The covariance matrix $R_{ij}$ can also be used to calculate an expectation value for the magnitude of the bulk flow, a convenient number with which to compare different spectra and catalogs. First, we diagonalize $R_{ij}$ to find the lengths of the axes of the covariance ellipsoid, $\sigma_1$, $\sigma_2$, and $\sigma_3$. The theoretical expectation value for the magnitude of the bulk flow in a given sample is then given by

$$\Lambda = \frac{(\sigma_1\sigma_2\sigma_3)^{-1}}{(2\pi)^{3/2}} \int |\vec{V}| \exp\left(-\sum_i V_i^2/2\sigma_i^2\right) d^3V.$$

We also calculate the expectation value for the 'noise-free' case, $\Lambda^{(v)}$, by using $R_{ij}^{(v)}$ instead of $R_{ij}$.

To study the likelihood that a given power spectrum could have produced both the LP and RPK results, we construct a 6-dimensional vector $\vec{U}^T = (\vec{U}^{\text{LP}}; \vec{U}^{\text{RPK}})$. Using a similar analysis to that described above, we calculate a covariance matrix $R_{ij}^T \equiv \langle U_i^T U_j^T \rangle$ and a corresponding $\chi^2$ for 6 degrees of freedom given by $\chi_T^2 \equiv U_i^T R_{ij}^T U_j^T$.

We consider power spectra from the IRAS–QDOT survey (Feldman, Kaiser & Peacock 1994), the BBKS standard CDM ($\sigma_8 = 1$, $\Omega h = 0.5$) model (Bardeen *et al.* 1986), a CDM spectrum using the maximum likelihood parameters calculated by Jaffe & Kaiser (1995) for the LP survey ($\Omega h = 0.075$, $\sigma_8 = 0.9$), CHDM simulations (normalized to COBE $Q_2 = 17\mu$K, Klypin *et al.* 1993), and PIB generated power spectrum ( $\Omega = 0.1$, $\lambda = 0.9$, Peebles 1994).

In Figure 1 a we show the power spectra we use to analyze and compare the data.

In Figure 1 b we show the trace of the squared tensor window function for the RPK and LP samples. For comparison we show the window function for the IRAS–QDOT sample (1824 galaxies) which probes a similar volume. From this figure it is clear that, except on the largest scales, the RPK and LP samples probe the power spectrum in very different ways. This implies that while both vectors will have similar contributions from the very largest scales, contributions from smaller scales will in general not be correlated.

In Figure 1 c we show the trace of the squared tensor window function for mock surveys with the same distribution in redshift as the RPK sample. As the sample size increases, the small scale contributions decrease.

## III. Results

In table 1, we show $\chi^2$ for the RPK result, the LP result, and both results taken together using a variety of power spectra. We also include the measure $\mathcal{C}$ of the expected correlation. Note that the RPK result is quite consistent with all the power spectra we have considered.

If there were no overlap between the RPK and LP window function, then the resulting bulk flow vectors would be expected to be uncorrelated and $\chi_T^2$ would be the sum of $\chi_{RPK}^2$ and $\chi_{LP}^2$. Window function overlap gives cross–terms which tend to favor agreement between the two vectors; *i.e.* if the window functions are similar then the vectors should be too. Here, since the RPK and LP bulk flow vectors point in almost opposite directions, overlap will increase $\chi_T^2$ so that the probability of getting both vectors decreases. Power spectra with lots of power on large scales, where the overlap is greatest, will be more strongly disfavored due to the higher expectation for correlation between the two results. As we see from table 1, this effect is greatest for the $CDM_{ML}$ spectrum (which has the most power on large scales, see Figure 1). However, even here the probability that both the RPK and LP results could be obtained is not small. For the other spectra, the large value of $\chi_T^2$ can be attributed to the large value of $\chi_{LP}^2$; indeed, the inclusion of the RPK data increases the likelihood for the IRAS–QDOT, CDM, and CHDM spectra.

Table 1: $\chi^2$ for RPK, LP and Total

|  | QDOT | CDM | CHDM | $CDM_{ML}$ | PIB |
|---|---|---|---|---|---|
| $\chi_{LP}^2$ | 10.40 | 10.41 | 10.43 | 5.52 | 11.33 |
| $P(\chi^2 > \chi_{LP}^2)$ | 0.015 | 0.015 | 0.015 | 0.137 | 0.010 |
| $\chi_{RPK}^2$ | 2.82 | 2.58 | 2.73 | 1.28 | 3.43 |
| $P(\chi^2 > \chi_{RPK}^2)$ | 0.420 | 0.461 | 0.435 | 0.734 | 0.323 |
| $\chi_T^2$ | 14.11 | 13.81 | 14.11 | 10.84 | 16.81 |
| $P(\chi^2 > \chi_T^2)$ | 0.028 | 0.032 | 0.028 | 0.093 | 0.010 |
| $\mathcal{C}$ | 0.08 | 0.07 | 0.08 | 0.35 | 0.11 |

*The $\chi^2$ and Probability for the LP and RPK surveys and the total quantities. Note that the quantity $P(\chi^2 > \chi_T^2)$ is calculated for six degrees of freedom whereas $P(\chi^2 > \chi_{LP}^2)$ and $P(\chi^2 > \chi_{RPK}^2)$ are calculated for three degrees of freedom. Also, the correlations of the two vectors are shown.*

Comparison of the results of the RPK and LP studies assumes that they are measuring the same quantity. However, an examination of Figure 1b and Table 1 shows that this is not necessarily the case. The RPK and LP bulk flow vectors contain significant contributions due to noise and incomplete cancelation of small scale flows. Both of these contributions depend on the details of the survey and of the power spectrum and would not be expected to correlate across different samples. As we have discussed above, the effect of the disagreement between the RPK and LP results is to disfavor models with large amounts of power on large scales, although not at a level that provides significant constraints.

Clearly, if RPK type measurements are to constrain the power spectrum, many more objects will be needed. If we assume that the power spectrum on large scales is not too far different from those considered in this Letter (excluding $CDM_{ML}$), then from

Figure 2 we can estimate that the signal to noise should become $\approx 1$ when the number of SN Ia's in the sample is of order 100. When the number of objects reaches 200, one should have a fairly precise value for the bulk flow of the sample. Given the greater sensitivity in the $z$ direction, it is likely that the SN Ia estimate for the $z$ component of the bulk flow could be reasonably accurate with just 60 or so objects. Between the Calán/Tololo survey and the CfA collection of supernovae, the RPK sample size should reach 40 by the end of 1995. However, the ending of the Calán/Tololo search effort makes the prospects dim for significantly increasing this number in the near future.

In contrast, for an LP type survey, with typical distance errors of approximately 15%, a sample size of the order of 300 is needed to get a a signal to noise of about one for our power spectra, or about 200 data points for the $z$ component $V_z$. Of course, if the actual bulk velocity is larger than the expectation values we have calculated, then it will show in a sparser survey; indeed, LP may have already detected a large $z$ component for the bulk flow.

Lauer and Postman in collaboration with Strauss have begun to take data for a survey similar to that of LP, but deeper ($R_{\max} = 24,000 \text{km/s}$). In addition to the BCG distance indicators that were used in the LP survey, they will use some velocity dispersion measurements that may tighten the errors down to about 10% of the distance. The survey will have about 600 clusters. To get the noise to be below the signal for a volume limited sample of this depth one would need at least 1200 data points for 15% error, (800 for a 10% error). However, to resolve the $z$ component, 800 points (550 for 10% error) would suffice. Thus, if the bulk flow is indeed large, or the error in the distance indicators is reduced to $< 10\%$, then the Lauer, Postman and Strauss survey may indeed see it. Given that the number density of objects in these surveys cannot be increased (it is a fairly complete sample), the only thing that can be done to improve the resolution of the bulk flow vector is to decrease the error in the distance measurements.

In Figure 3 we show the effect of reducing the errors on an LP type survey. Here S/N becomes 1 when the errors are $\sim 5\%$, for the $Z$ direction it is enough to have errors of $\sim 8\%$.

**Acknowlegements:** This research was supported in part by the National Science Foundation and by NASA grants NAG5–1310 and NAG5–2412.

# Figure Captions

**Figure 1 a** *The redshift corrected power spectra used in the analysis.*

**Figure 1 b** *The trace of the squared tensor window functions for the RPK survey, as well as the LP ones. For comparison we show the window function for the IRAS–QDOT sample (1824 galaxies) which probes a similar volume. It is clear that, except on the largest scales, the RPK and LP samples probe the power spectrum in very different ways. This implies that while both vectors will have similar contributions from the very largest scales, contributions from smaller scales will in general not be correlated.*

**Figure 1 c** *Window functions for RPK type surveys of different sizes. The contributions from large k fall as the number of data points increase.*

**Figure 2** *The noise–free expectation values for the z component of $\Lambda^{(v)}$ and its magnitude for the four power spectra considered as a function of the size of the survey for an RPK–type sample. We also show the expected magnitude of the noise, $\Lambda^{(\varepsilon)}$, which falls with the number of data points.*

**Figure 3** *The noise–free expectation values for the z component of $\Lambda^{(v)}$ and its magnitude for the four power spectra considered as a function of the measurement errors for an LP–type survey. We also show the expected magnitude of the noise, $\Lambda^{(\varepsilon)}$, which increases with increasing measurement errors.*

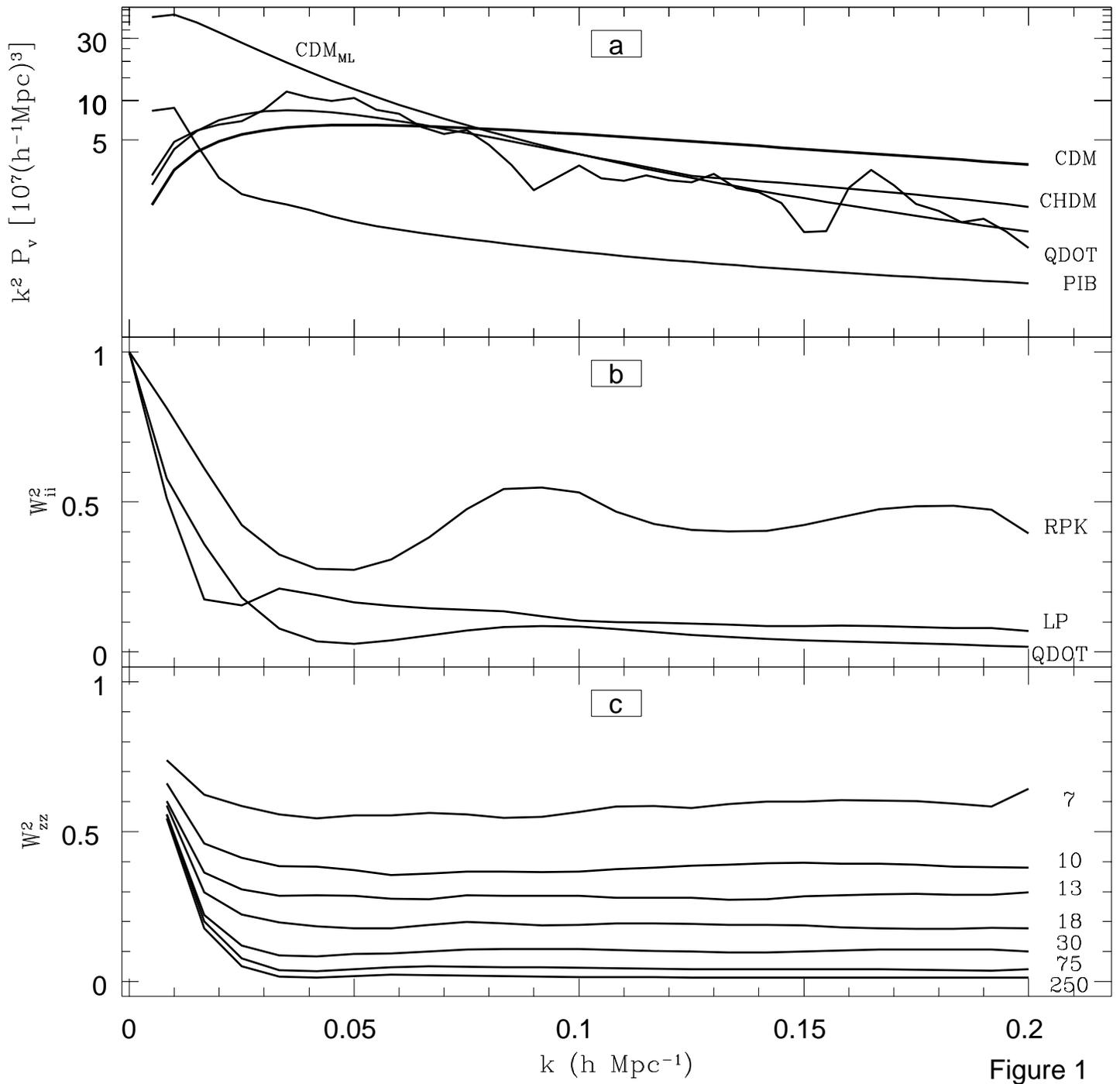

Figure 1

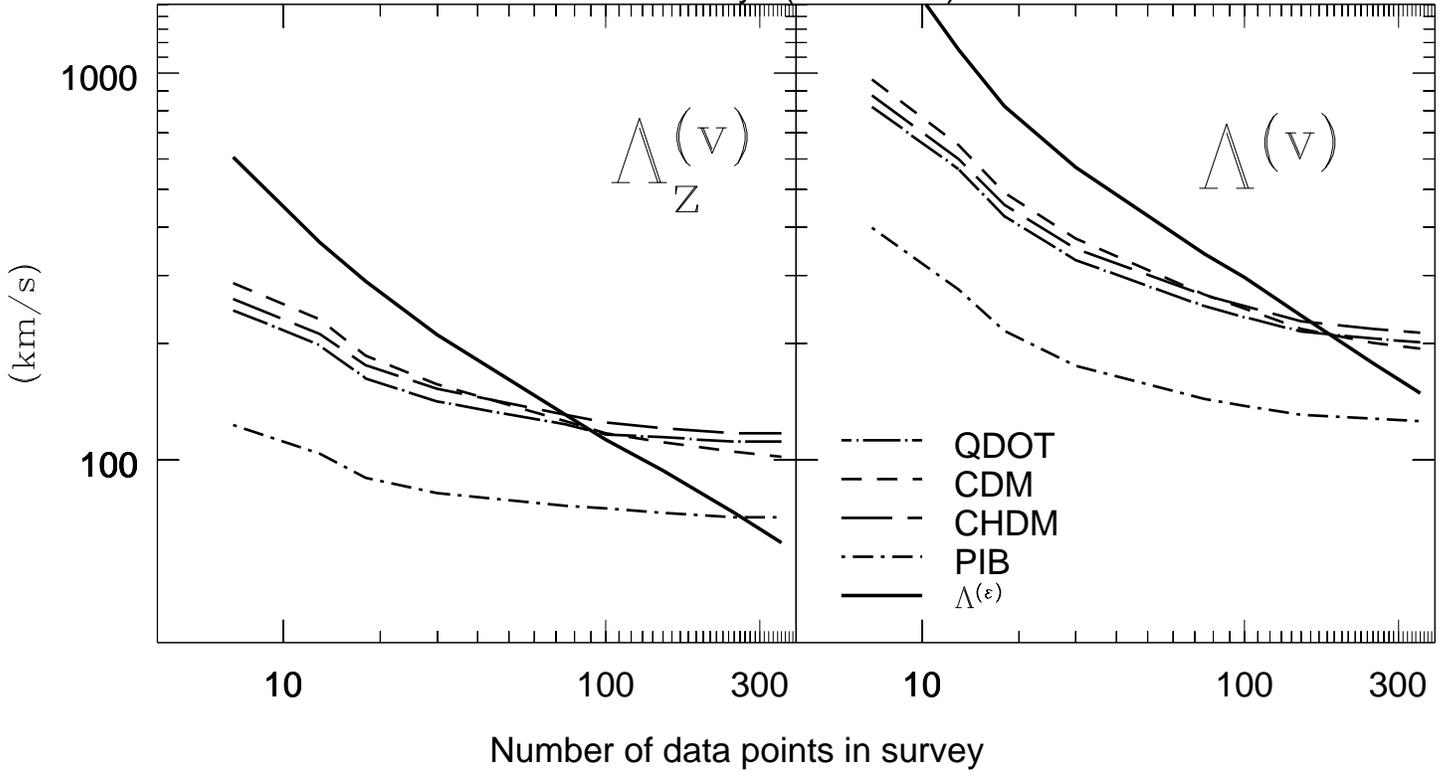

Figure 2

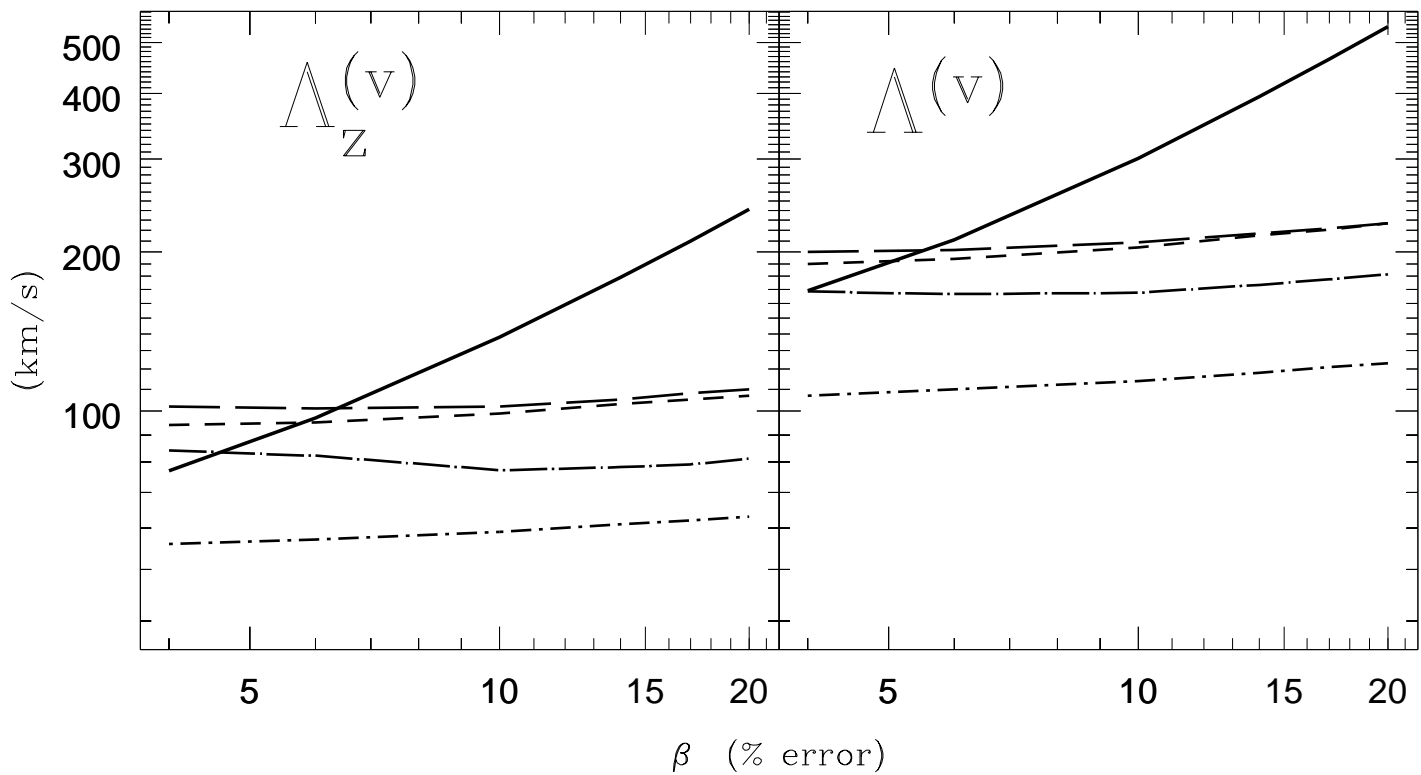

Figure 3